\documentclass[11pt, twocolumn]{article}

\usepackage[margin=0.95in]{geometry}
\usepackage[T1]{fontenc}
\usepackage[utf8]{inputenc}
\usepackage{lmodern}
\usepackage{microtype}
\usepackage{csquotes}
\usepackage{enumitem}
\usepackage{booktabs}
\usepackage{amsmath, amssymb}
\usepackage{titlesec}
\usepackage[hang]{footmisc}
\usepackage[round,authoryear]{natbib}

\usepackage{hyperref}

\expandafter\def\expandafter\UrlBreaks\expandafter{\UrlBreaks%
  \do\a\do\b\do\c\do\d\do\e\do\f\do\g\do\h\do\i\do\j%
  \do\k\do\l\do\m\do\n\do\o\do\p\do\q\do\r\do\s\do\t%
  \do\u\do\v\do\w\do\x\do\y\do\z\do\A\do\B\do\C\do\D%
  \do\E\do\F\do\G\do\H\do\I\do\J\do\K\do\L\do\M\do\N%
  \do\O\do\P\do\Q\do\R\do\S\do\T\do\U\do\V\do\W\do\X%
  \do\Y\do\Z\do\0\do\1\do\2\do\3\do\4\do\5\do\6\do\7\do\8\do\9\do\.}

\hypersetup{
  colorlinks=true,
  breaklinks=true,
  linkcolor=blue,
  citecolor=blue,
  urlcolor=blue,
  pdftitle={Agentic Inequality},
  pdfauthor={Matthew Sharp},
  pdfcreator={Matthew Sharp},
  pdfproducer={Matthew Sharp},
  pdfkeywords={AI agents, inequality, governance, agentic inequality}
}
\usepackage{bookmark}

\titleformat{\section}{\large\bfseries}{\thesection.}{0.5em}{}
\titleformat{\subsection}{\normalsize\bfseries}{\thesubsection.}{0.5em}{}
\titleformat{\subsubsection}{\normalsize\itshape}{\thesubsubsection}{0.5em}{}
\titlespacing*{\section}{0pt}{3.5ex plus 1ex minus .2ex}{2.3ex plus .2ex}

\title{\textbf{Agentic Inequality}}
\author{
  Matthew Sharp\textsuperscript{1,*} \and Omer Bilgin\textsuperscript{2} \and
  Iason Gabriel\textsuperscript{3} \and Lewis Hammond\textsuperscript{2,4}
}
\date{28 August 2026}
\newcommand{\affiliations}{%
  \vspace{-1em}
  \begin{center}
    \footnotesize
    \textsuperscript{1}Oxford Martin AI Governance Initiative \quad
    \textsuperscript{2}University of Oxford \\
    \textsuperscript{3}Google DeepMind \quad
    \textsuperscript{4}Cooperative AI Foundation
  \end{center}
  \vspace{0.5em}
}
\begin{document}

\twocolumn[
  \begin{@twocolumnfalse}
    \maketitle
    \affiliations
    \begin{abstract}
      \noindent As autonomous AI agents capable of complex planning and action are deployed more widely, questions about who can access them, how capable they are, and how many of them users can deploy are likely to shape the distribution of power and outcomes at a societal level. We characterise the resulting disparities as \textit{agentic inequality}. We show that agents could either deepen existing divides or, under the right conditions, mitigate them. The paper makes three contributions. Firstly, it develops a framework for analysing agentic inequality across three dimensions – \textit{availability}, \textit{quality}, and \textit{quantity}. Secondly, it argues that agentic inequality is not the same as earlier technological divides because AI agents function as autonomous delegates (rather than tools) and so generate new asymmetries through scalable goal delegation and direct agent-to-agent competition. Thirdly, it analyses the technical, socioeconomic, and political drivers of agentic inequality, such as model release strategies and market incentives, that are likely to shape the distribution of agentic power. We conclude by setting out a research agenda for governance.
      \vspace{2em}
    \end{abstract}
  \end{@twocolumnfalse}
]

\renewcommand{\thefootnote}{}\footnotetext{\raggedright\textsuperscript{*}\textit{Corresponding author:} \url{matthew.j.sharp@gmail.com}.}\renewcommand{\thefootnote}{\arabic{footnote}}

\section{Introduction}

Generative AI models, typified by systems that convert prompts into prose, code or images, now underpin AI agents – systems that pursue complex goals with significant autonomy, which involves perceiving their environment, planning their actions and then executing multistep tasks across digital environments \citep{ref01}. These systems vary in autonomy, efficacy, goal complexity, and generality \citep{ref02}. For example, they range from narrow task executors operating within constrained workflows to general-purpose delegates managing cross-domain goals. AI agents are still at an early stage of development and deployment. However, they are beginning to perform economically valuable work and ongoing research aims to further extend their capabilities. This shift from AI as a command-following tool to AI as an autonomous, goal-pursuing actor matters because it affects human agency: the capacity of individuals to pursue the goals they value in the way they desire \citep{ref03,ref04}.

This paper introduces and explores \textit{agentic inequality}, which we define as the disparities in power and outcomes that may arise from unequal access to AI agents, and from differences in their quality and number. We argue that agentic inequality differs from earlier technological divides. This is because AI agents can create new power asymmetries by allowing users to delegate complex tasks at scale and new competitive dynamics through direct agent-to-agent competition. Our argument sits within broader political economy debates about the use of AI, but it focuses on disparities arising from the distinctive features of AI agents. Since these systems are not yet widely deployed, a window of opportunity exists to proactively shape their development and ensure their benefits are equitably distributed.

We build on scholarship on the co-evolution of technology and society \citep{ref05,ref06} and extend research on large language model (LLM) impacts \citep{ref07,ref08,ref09} and the “digital divide” \citep{ref10,ref11,ref12}. We develop an analytical framework for this emerging challenge that explores the risks of new forms of inequality as well as the potential for agents to mitigate existing divides, while being careful not to present any outcomes as inevitable. Agents could amplify the advantages of well-resourced actors, but if capable systems are widely accessible, they could also reduce existing disadvantages by simplifying complex processes and performing tasks that previously required costly expertise. To understand this distributive effects more fully, we need to view AI agents not only as individual entities but also as comprising a form of infrastructure that may mediate access to essential goods and services as well as opportunities \citep{ref13,ref14,ref15,ref16,ref17}. Our focus is primarily on the use of agents by individual people and organisations, although similar dynamics extend to competition between states.

The paper proceeds as follows. Section 2 identifies the key dimensions through which agentic inequality may manifest. Section 3 examines its implications across key societal domains. Section 4 analyses the technical, socioeconomic and political drivers that may widen or reduce agentic inequality. Section 5 considers the governance challenge and sets out a research agenda to tackle this.

\section{Dimensions of Agentic Inequality}

Agentic inequality has three dimensions: \textit{availability}, \textit{quality}, and \textit{quantity}. Each captures a distinct source of differential advantage, and together they determine how much value an actor can derive from agentic AI.

\subsection{Availability}

\textit{Availability} is the most basic dimension of agentic inequality and it has two layers. \textit{Structural availability} concerns whether an AI agent is available to a given actor, taking into account commercial availability, regulatory permissibility, and the necessary infrastructure to run it. \textit{Accessibility} concerns whether an actor can use an agent despite constraints around affordability, digital literacy, language support, usability, and so on. This distinction reflects the fact that the formal access, entailed by structural availability, does not automatically mean effective use, a distinction that has been noted in research on the digital divide for some time \citep{ref10,ref11,ref12}. Together, these two conditions determine whether an individual or organisation can use even a single AI agent, which is the minimum requirement for participation in the agentic economy \citep{ref18}.

A further distinction is important here: access to a foundation model is not the same as access to a functional agent. Foundation models are becoming more accessible, but at the time of writing, it is difficult to use capable agents because essential infrastructure, including high-quality scaffolding, is still new, complex, and concentrated among just a few providers. Specialist expertise is also needed for effective deployment \citep{ref16,ref19}, so knowledge gaps are a barrier to entry too. Not all use disparities reflect barriers, though: sometimes they reflect a choice made by an individual or organisation not to use agents for ethical or practical reasons.

\subsection{Quality}

\textit{Quality} refers to an individual AI agent’s capabilities and the operational characteristics that determine its behaviour. Agent quality can vary along several axes:

\begin{itemize}
    \item \textit{Core intelligence:} the combination of an agent’s world knowledge, its multimodal capabilities across language and other data types, and its ability to reason, plan, and solve problems. This depends largely on how sophisticated the underlying foundation model and its training data are \citep{ref20}.
    \item \textit{Operational speed and throughput:} an agent’s efficiency depends on latency and data-processing capacity, which in turn reflect the computational resources that support the agent, including hardware accelerators and cloud infrastructure \citep{ref21}.
    \item \textit{Reliability:} the consistency and robustness of an agent’s performance across tasks.
    \item \textit{Tool use:} an agent’s ability to access and effectively use external systems, such as APIs, proprietary databases, real-time data feeds, or physical actuators \citep{ref16}.
    \item \textit{Disposition:} an agent’s behavioural tendencies. These affect performance in context-specific ways. For example, an aggressive disposition can help in negotiation but hinder performance in a customer service setting where politeness and helpfulness are more important.
\end{itemize}

These subcomponents need not move together, meaning that quality gaps may shrink on one axis while remaining wide on another. For example, even when tool use remains gated by infrastructure controllers, falling inference costs \citep{ref22} may make core intelligence cheaper. Among these different axes, \textit{core intelligence} and \textit{tool use} are likely to matter most for inequality. The former determines an agent’s ability to perform complex, high-value tasks; the latter determines whether an agent can act in the world and not just advise. \textit{Disposition} may also become increasingly important as multi-agent interactions become commonplace, since behavioural tendencies in competitive or cooperative settings will directly shape outcomes.

\subsection{Quantity}

The \textit{quantity} of agents an individual or organisation can use captures a source of advantage distinct from the capability of any single agent. Scale can matter because running more agents means that a task can be pursued through multiple paths in parallel rather than sequentially. Recent research finds that performance on complex tasks can improve as more agents are deployed, with larger gains on harder tasks \citep{ref23}, and other work finds that multi-agent teams running in parallel can reduce latency while preserving accuracy, and in some cases improve task completion rates \citep{ref24}. Quantity may therefore confer an advantage not just by adding effort, but also by enabling broader search and faster convergence across complex problems.

The dimensions of quantity and quality may reinforce or substitute for one another. Greater quantity can compensate for lower quality: several competent (though individually limited) agents may outperform a single more advanced agent, particularly when they draw on diverse models, prompts, tools, or assigned functions \citep{ref23,ref25}. Multiplying simple agents does not, however, guarantee better results: when agents are highly alike, their outputs increasingly overlap, and returns may diminish sharply \citep{ref26,ref27}. Where diversity and coordination improve at the same time, though, gains can compound: heterogeneous multi-agent systems can outperform single-agent baselines and larger but less differentiated groups of agents \citep{ref25,ref26}.

User outcomes also depend on complementary factors such as how far agents are integrated into workflows, how their use is overseen, what domain knowledge the user brings, and whether the user is able to use agents strategically \citep{ref08,ref28}. Even if access is equal in formal terms, realised benefits can be substantially different because the complementary skills and knowledge that turn agents’ outputs into value are unevenly held.

A related issue is \textit{user dependence}: how far people come to rely on agents for tasks they previously handled themselves. Dependence is unlikely to be spread evenly, as people with fewer alternative resources are likely to rely more heavily on agents, which exposes them more to quality issues or service interruptions \citep{ref04}. We therefore treat dependence not as a separate fourth dimension, but rather as a factor that can increase the significance of the dimensions described above. Dependence is not inevitable, however, as agents designed to explain their reasoning and build user competence over time could \textit{reduce} reliance.

Taken together, these dimensions help illuminate two features that distinguish agentic inequality from earlier technological divides. The first is scalable delegation: using many agents in parallel is more akin to directing a distributed workforce than it is to using a better tool. The second concerns direct competition between agents, which themselves may negotiate, bid, plan, and strategise on behalf of users. As a result, contests that once rested on how people choose to interact, and the choices they make, may instead be decided by networks of agents that operate at machine speed \citep{ref20}.

\section{Implications of Agentic Inequality}

Agentic inequality could have substantial economic, social, and political effects. The ways in which differences in the \textit{availability}, \textit{quality}, and \textit{quantity} of agents unfold across domains, as well as the subsequent policy choices that are made, will determine whether agents concentrate power among actors who are already well-resourced or create wider access to opportunity. The discussion here draws, where possible, on existing evidence and, where the technology is still emerging, projects from present trends. In some places, it also considers longer-term possibilities.

\subsection{Economic Impacts}

The \textit{labour market effects} of agentic AI remain uncertain. \textit{Agentic capital}, or autonomous systems that can manage entire business processes, has the potential to accelerate a shift in national income share from labour to capital, extending a dynamic observed with certain previous technological waves \citep{ref07,ref29}. However, evidence on generative AI’s labour market impact thus far is mixed, and most of it predates autonomous agents. Economy-wide analyses suggest relative stability \citep{ref30}, but disaggregated studies of exposed industries find that hiring of junior roles has declined while employment for senior workers has remained stable or grown \citep{ref31,ref32}. Some firm-level studies demonstrate a levelling-up effect, with technology raising the productivity of staff with less experience \citep{ref08,ref09,ref33,ref34}, but this effect is found in the context of human–tool augmentation – it is far from clear that it would survive a transition to agentic capital, which is built to execute and manage entire workflows with limited human input.

\textit{Industrial organisation and market structures} could also be affected by agentic inequality. The central question is whether agents intensify concentration or widen competition. On one hand, the rise of superstar firms could be accelerated through two mechanisms. Firstly, dominant firms could use proprietary data to drive more capable agents; those agents could also allow them to improve services and attract more users, and to generate further data in a feedback loop that smaller rivals may be unable to match \citep{ref35}. Secondly, larger firms could also use more agents to automate internal processes and accelerate innovation at a scale beyond the reach of smaller competitors. On the other hand, if powerful agents become widely available and the price of model inference continues to fall \citep{ref22}, the reverse could occur: agents could lower entry barriers and startups could run operations that previously required substantial human capital.

Looking next to the domain of \textit{consumer welfare and negotiations}, corporate agents could amplify manipulative practices – including deceptive choice architecture, scarcity cues, and hidden information – already documented in digital markets, at a scale and speed that no human sales operation can match \citep{ref36,ref37}. Differences in agentic capabilities could lead directly to worse outcomes for the weaker party, whether in negotiations between consumers and firms or between differently resourced firms \citep{ref20}. Pricing algorithms could also weaken competitive pressure by learning tacitly collusive strategies that keep prices above competitive levels \citep{ref38}. These practices could be counteracted, however, through widely available consumer agents automating comparison shopping, flagging deceptive pricing, and negotiating on users’ behalf \citep{ref20,ref39}.

\subsection{Social and Political Impacts}

Agents could reshape how citizens interact with the state, producing either a more stratified system or wider \textit{access to public services}\citep{ref40}. Affluent individuals could use premium agents, able to manage complex bureaucratic workflows, to secure better outcomes in, for example, healthcare or legal systems \citep{ref41}. If public services start being designed around agentic interaction, citizens without agents risk being disadvantaged. Managed well, however, public-good agents could widen access by automating form-filling and proactively pursuing entitlements, thereby reducing administrative burdens that disproportionately affect lower-income groups. Taiwan’s experience with digital public infrastructure is one instructive precedent of this \citep{ref42}.

In terms of \textit{political discourse and participation}, agents could either amplify elite influence or enable broader participation \citep{ref43}. Better-resourced actors could deploy sophisticated agent swarms to run coordinated influence campaigns, overwhelm public consultations, or generate personalised propaganda on a large scale, highlighting again how gaps in agent quantity and quality could allow more influence \citep{ref44,ref45}. At the same time, broad access to agents that can autonomously draft policy proposals, manage activist campaigns, and engage in political processes could strengthen the capacity of citizens and grassroots groups. Recent work has begun to explore how AI might support participatory deliberation in digital public spaces at scale \citep{ref46}. This shift from human persuasion to machine-mediated influence also raises a deeper question for democratic legitimacy, since the “public will” may come to be shaped more heavily by automated political action.

Widely accessible agents could improve social mobility by providing coaching and opening opportunities that might otherwise remain out of reach. They could also reduce the transaction costs that inhibit beneficial coordination and enable new forms of decentralised social bargaining \citep{ref47}. If unequally distributed, however, AI agents could lead to greater \textit{social stratification}. For example, agents could deepen the Matthew Effect, whereby initial advantages compound over time, widening gaps in social capital. They also allow complex social strategies to be delegated at scale, such as managing professional networks or improving educational opportunities for one’s children.

\section{Forces Shaping Agentic Inequality}

The determinants of agentic inequality include a mix of technical constraints, market incentives, institutional arrangements, and political choices. Many of these factors apply to AI more broadly, but they matter more for agents, since systems that act autonomously on behalf of their users have more direct impacts on distributive outcomes. This section focuses on technical, socioeconomic, and political drivers of agentic inequality. These drivers may interact with broader structural inequalities, including those related to class, race, gender, disability, and geography. While we do not attempt to systematically trace such interactions in this paper, they warrant further interdisciplinary attention.

\subsection{Supply-Side and Ecosystem Drivers}

\textit{Compute costs and capital barriers} are the most immediate constraints: very large capital investments are needed to train foundation models, which only a few major technology firms can afford \citep{ref21,ref48}. This concentration affects both the availability of the most powerful AI agents and the level of core intelligence they provide. Inference costs, too, although much lower than model training costs, can become expensive as tasks get more complex and require more tokens to complete. More capable models also typically charge higher per-token prices. These dynamics lead to a pay-for-performance pattern, in which premium tiers supply stronger reasoning or faster responses. Where those gains taper off, however, a good-enough tier may prove viable for most baseline autonomous uses, even if a quality advantage is retained by frontier systems.

Other areas that affect agentic inequality are \textit{agent architecture and platform governance}. How foundation models are governed – whether proprietary or open-weight – shapes how agentic capabilities spread and who retains control over them \citep{ref49}, though neither governance model is inherently more equitable – each comes with its own patterns of access constraints, dependency risks, and quality differences \citep{ref50}. Releasing weights openly lifts one constraint, since the model itself is in the user’s hands. Compute, engineering capacity, and implementation expertise then become binding inputs, and better-resourced actors have an advantage. Hosted access via APIs, the standard route to proprietary models, lowers a significant barrier: the user avoids the expense and operational burden of running the infrastructure, though this convenience could result in dependence. Users may not be able to tailor model weights to private data, may have limited control over updates, and may rely on API access controlled by the provider. These limits can affect whether a downstream agent carries out complex tasks consistently.

An agent’s value depends heavily on its connection to APIs, proprietary databases, and large software systems – thus \textit{agentic integration into the wider digital environment} is another influence on agentic inequality. Actors that control this \textit{agent infrastructure}, including cloud providers, software platforms, and financial institutions, are therefore significant gatekeepers to agentic capabilities. Through API policies, data-sharing terms, and platform rules, they can widen or narrow benefits for different users and developers. These levers directly affect agent quality, since effective tool use is central to performance \citep{ref14,ref16}. While open standards and interoperability requirements can soften the impacts, these chokepoints can create new forms of dependence.

\subsection{Sociopolitical and Institutional Forces}

\textit{Economic incentives and market dynamics} can work in both directions. For example, AI companies can build up their user bases via cheap offers of basic agents. At the same time, firms optimising for profits via product differentiation and price discrimination can increase inequalities by introducing gaps in agent quality and quantity – if a customer can afford a premium tier, they can get better planning, more reliable performance, or a wider set of tools. Higher pricing for large agent swarms may price individuals and smaller organisations out of scalable delegation functions. Generally, the picture is likely to be one of broad access to basic systems but continued inequality in the most valuable uses.

\textit{Digital literacy and human–agent interaction} are likely to matter for effective use of AI agents. In the short run, setting complex goals and managing multistep agent delegation may require high levels of digital literacy. Specialised technical skill might be expected to lose significance over time, however, as interfaces become more intuitive, for example by allowing users to use natural language to delegate increasingly complex tasks \citep{ref44}. An analogous case is ChatGPT adoption, where its language-based interface has supported rapid growth in low- and middle-income countries \citep{ref51}, far beyond what was seen with earlier, more technical AI tools.

In terms of \textit{geopolitics and jurisdictional fragmentation}, agentic inequality can surface at the global level where techno-national competition and the need for international coordination are out of balance. Governments are already designing protectionist policies in response to national security concerns, including semiconductor export controls that restrict access to foundational AI hardware \citep{ref52,ref53}. The same could happen with software as agents become more sophisticated: states pursuing mercantilist strategies could produce export versions of agents whose planning abilities are weaker, or whose autonomy is more limited, compared to top-tier domestic options. In addition to this embedding of agentic power asymmetries internationally, jurisdictional fragmentation can worsen the problem at the level of firms: divergent regulations on issues such as data privacy across countries can create compliance burdens that large, multinational firms are more able to absorb than smaller competitors \citep{ref54}. An emerging international consensus around AI safety and responsible use, however, could provide at least some basis for more coordinated and equitable governance \citep{ref55,ref56}.

Any response to agentic inequality is constrained by the pacing problem: technological change may advance much faster than institutions can adapt \citep{ref57}. The resulting lag entrenches inequalities in the ability to deploy AI agents, especially where regulatory capture is also a risk \citep{ref58}, though it also leaves scope for anticipatory governance. The difficulty here is to design institutions that are capable of shaping how agents are used so that unequal patterns do not calcify into permanent structures \citep{ref59,ref60}.

\section{Governing Agentic Inequality: Complexity and Future Directions}

Governance should not only prevent the harms of disparity but create the conditions in which agents can function as an equalising force. This is a difficult problem because it begins with a normative question: what distribution of agentic capabilities should count as fair \citep{ref61}? Societies already tolerate large disparities in access to expert human services, such as legal or financial advice. Should inequalities in agentic capability be judged by the same standards as disparities in human capability, or do agents’ scalability and autonomy demand stricter standards of justice?

A full account of agentic equality – of what a fair distribution of agentic capabilities would require – is beyond the scope of this paper. However, two directions seem promising: ensuring universal structural availability as a threshold condition, especially when AI agents function as a gateway to service-access and opportunity, and designing agent infrastructure with public-interest considerations in mind.

Substantial practical obstacles will remain even if some normative consensus emerges. Existing legal frameworks are not well-suited to harms produced by unequal agentic capabilities, because disadvantages can arise through cumulative asymmetries rather than a single, clearly defined wrong \citep{ref37}. Attempts to level the playing field through direct intervention also face serious coordination problems and encounter the Collingridge dilemma: acting early is difficult because uncertainty is high, while acting late risks the technology already being entrenched \citep{ref62}. For example, caps on agent power could constrain innovation, while mandates to redistribute compute could divert resources from large-scale projects that have substantial benefits to society. Government efforts to guarantee universal access through a single provider could produce vendor lock-in and weaken competition, thereby reinforcing rather than reducing inequality \citep{ref21,ref58}.

The path forward requires a \textbf{research agenda} that cuts across the three core dimensions of \textit{availability}, \textit{quality}\textit{,} and \textit{quantity}, to build the knowledge needed to govern disparities around each. Progress across these programmes will necessitate sustained interdisciplinary engagement, connecting the framework developed here to scholarship on structural inequity \citep{ref10,ref11,ref12} and regulatory governance \citep{ref57,ref58,ref59}.

A first priority is \textit{setting the empirical foundations}. Before effective interventions can be designed, researchers need robust metrics to track disparities in agent \textit{availability}, \textit{quality}, and \textit{quantity} in real-world settings. Many of the technologies likely to drive this inequality, such as automated negotiation or scaled task delegation, remain novel and largely unmeasured. Key questions include: What are the best methods for empirically tracking the deployment of AI agents and their distributional impacts? How can we carry out direct technical comparisons to assess how far disparities in agent quality allow more capable agents to persuade or manipulate weaker ones? Possible indicators include agent deployment rates by income decile and geography; benchmark performance gaps between free and premium agents on standardised task suites; and performance differentials in controlled strategic interactions across different capability tiers.

A second programme is about \textit{establishing} \textit{the normative foundations} that help to define fairness across the three dimensions. Once empirical patterns are clearer, a key task is determining which disparities in agent \textit{availability}, \textit{quality}, and \textit{quantity} are unacceptable, when judged from a social or ethical standpoint. Key questions include: What kind of differences in agent capability matter, and under what conditions? Which participatory methods (such as citizens’ assemblies and deliberative polling) best elicit public preferences about the acceptability of capability gaps? How can these preferences be translated into concrete design principles or regulatory guardrails governing disparities in agent quality?

A third concern is designing the agentic ecosystem itself – specifically the development of \textit{technical and infrastructural levers} \textit{that can support} \textit{equality}. The aim is to build equity-supporting features directly into technical and infrastructural arrangements, reducing harmful capability gaps and limiting excessive concentration in the ability to use large numbers of agents. Key questions include: How should we govern critical resources such as compute in ways that balance innovation with equitable access? Which interoperability standards are needed to prevent platform lock-in from entrenching agent quality disparities \citep{ref16,ref63}?

A fourth research area is \textit{public service models for securing equity}. This includes policies intended to guarantee universal access and ensure a sufficient baseline for agent quality, recognising that government-led digital provision has a mixed track record \citep{ref64}. One promising idea is \textit{universal basic agency}, which sees access to AI agents as a primary good in Rawls’s sense and thus something all citizens are entitled to as a matter of justice \citep{ref65}. On this account, agents can be understood to function as all-purpose means, enabling people to pursue a wide range of ends and conceptions of the good life. Justice therefore requires more than formal availability alone. People must be able to access and use agents of sufficient quality for that capability to be genuinely effective across the population, even if others can deploy more powerful or more numerous systems. Key questions include: Should a government provide a public-option AI agent, or would subsidising private services be more effective? What level of agent quality would constitute a meaningful empowerment baseline? How can public agents, or public infrastructure for private agents, be designed to serve public interests rather than commercial ones? Early proposals for public AI infrastructure suggest that this is already a live policy question in several jurisdictions \citep{ref66,ref67}.

A final programme concerns the \textit{regulation of asymmetries in agent quality and quantity}. These new asymmetries strain existing legal doctrines, and the advantages conferred by superior agents may become entrenched without clearer rules. Key questions include: How can legal frameworks be adapted to provide recourse for individuals harmed by agent quality disparities \citep{ref68}? What agile or outcome-based regulatory models can keep pace with technical change while protecting people from exploitation by actors using vast numbers of agents \citep{ref57}?

\section{Conclusion}

Scalable delegation to AI agents creates a genuinely novel source of inequality, as does the shift from contests between people to contests between their agents. This new societal challenge is being driven by a range of technical, economic, and political forces. Many of the drivers of agentic inequality, however, are also potential levers for shaping more equitable outcomes, from model architecture to market design. Unfortunately, the window to address this issue will not remain open indefinitely: patterns of agentic use that become entrenched are far harder to reshape. The framework developed in this paper lays out the nature of this challenge, and the research agenda proposed above is an attempt to identify what needs to be understood before that point is reached.

\section*{Acknowledgements}

We thank Alan Chan, Noam Kolt, Seth Lazar, and Sam Manning for helpful feedback. This work was supported by the Future Impact Group and the Cooperative AI Foundation.

\section*{Competing interests}

I.G. is employed by Google DeepMind. M.S., O.B. and L.H. declare no competing interests.

\bibliographystyle{agsm}
\bibliography{Agentic_Inequality1}

\end{document}